# g$_r$enepipe: A flexible, scalable, and reproducible pipeline to automate variant and frequency calling from sequence reads

Lucas Czech[1,*], Moises Exposito-Alonso[1,2,*]

[1]Department of Plant Biology, Carnegie Institution for Science, Stanford, CA 94305, USA
[2]Department of Biology, Stanford University, Stanford, CA 94305, USA
[*]correspondence to: luc.czech@gmail.com, moisesexpositoalonso@gmail.com
ORCID IDs: 0000-0002-1340-9644, 0000-0001-5711-0700

## Abstract

**Processing high-throughput DNA sequencing data of individuals or populations requires stringing together independent software tools with many parameters, often leading to non-reproducible pipelines and datasets. We developed `g`$_r$`enepipe` to streamline this data processing, an all-in-one Snakemake workflow from raw sequencing data to the end product of a table of individuals' genotypes or population frequencies. Our pipeline allows users to select among a range of popular software tools within a single configuration file, automatically downloads and installs software and dependencies, and runs with two command calls: to prepare and to run. It is highly optimized for scalability in cluster environments and parallel computing, splitting data tasks into manageable genomic sections and automatically consolidating the outputs. `g`$_r$`enepipe` is published under the GPL-3 license, and freely available at https://github.com/moiexpositoalonsolab/grenepipe.**

## Introduction

High-throughput genomic sequencing technologies have revolutionized not only biomedical but also ecological and evolutionary research. Whether sequencing is conducted in DNA samples from single-cells of a tumor biopsy sample, or from pooled DNA of hundreds of individuals of a plant species growing in the wild, the core bioinformatic processing of sequencing reads is virtually the same: DNA sequencing reads (typically 30-250 letters long) are compared to a reference sequence or genome, and the variant positions (e.g., A→G, single nucleotide polymorphisms [SNPs], or AACG→ACG, insertions/deletions [indels]) are listed in a final variant call table. This process requires manipulating large datasets (GBs or TBs), and a myriad of processing steps that rely on different software tools. This bioinformatic challenge is often a limitation for biology trainees and research groups when adopting new sequencing technologies.

    We aimed to streamline the process of sequence data processing with `g`$_r$`enepipe`, an automated and flexible pipeline for variant and frequency calling from raw sequences. Although several workflows for such analyses exist in GitHub or elsewhere [1–4], these often require bioinformatic expertise, substantial pipeline preparation, or manual installation of tools. `g`$_r$`enepipe` does not require user inputs except for a table of sample names, and runs with a two-liner command.

    Its backend is the platform-independent Snakemake workflow engine [5]. This allows to automate large data processing by multiple tools, takes care of intermediate file bookkeeping and

execution order dependencies, and parallelizes independent tasks. All tools for individual steps are automatically loaded via Conda/Bioconda [6], the execution can seamlessly recover and continue after failed tasks, and workflows and results can be archived to facilitate reproducibility.

Our pipeline comprises all steps of the well-established GATK best-practices workflow [7], and adds recent popular tools and further quality controls with a focus on evolutionary ecology applications, including quality profiling of ancient DNA from historical specimens or old field samples [8], or within-library calculation of allele frequencies for Pool-Sequencing [9]. A high-level overview of the pipeline is shown in **Fig. 1**.

## Pipeline inputs and outputs

**Input**

The minimal input to the pipeline are one or more `fastq` files [10] representing distinct samples, e.g., sequenced by an Illumina sequencing instrument, either single-end or paired-end, and optionally gzip-compressed. In addition, a reference genome or sequence to compare with is required in `fasta` format [11]. Samples can also consist of multiple `fastq` files representing multiple units of the same sample, e.g. as a result of re-sequencing runs of the same biological sample. Optionally, if the user studies a species with an annotated genome in the SnpEff database [12], a suitable species name can be provided to also run SnpEff. In addition, if the user has a known set of genomic positions of interest, a reference `vcf` can also be provided for guided SNP calling.

**Output**

The main output of the pipeline is a variant call table (`vcf` file) [13], that contains single nucleotide (and short indel) variants across the genome in rows, for all input samples in columns. The second major output is a `MultiQC` quality control `html` report (including summaries of many intermediate tools). Optionally, a tab-separated allele frequency table, `frq`, of the ratios of bases within a sample can be computed for Pool-Seq approaches. The most relevant intermediate files (e. g., trimmed `fastq` files and mapped `sam`/`bam` files) are kept by default, and can hence be inspected or used for further downstream analyses. Lastly, the Snakemake backend can generate benchmarking and processing reports of the pipeline, which allows users to evaluate tool runtimes and resource requirements.

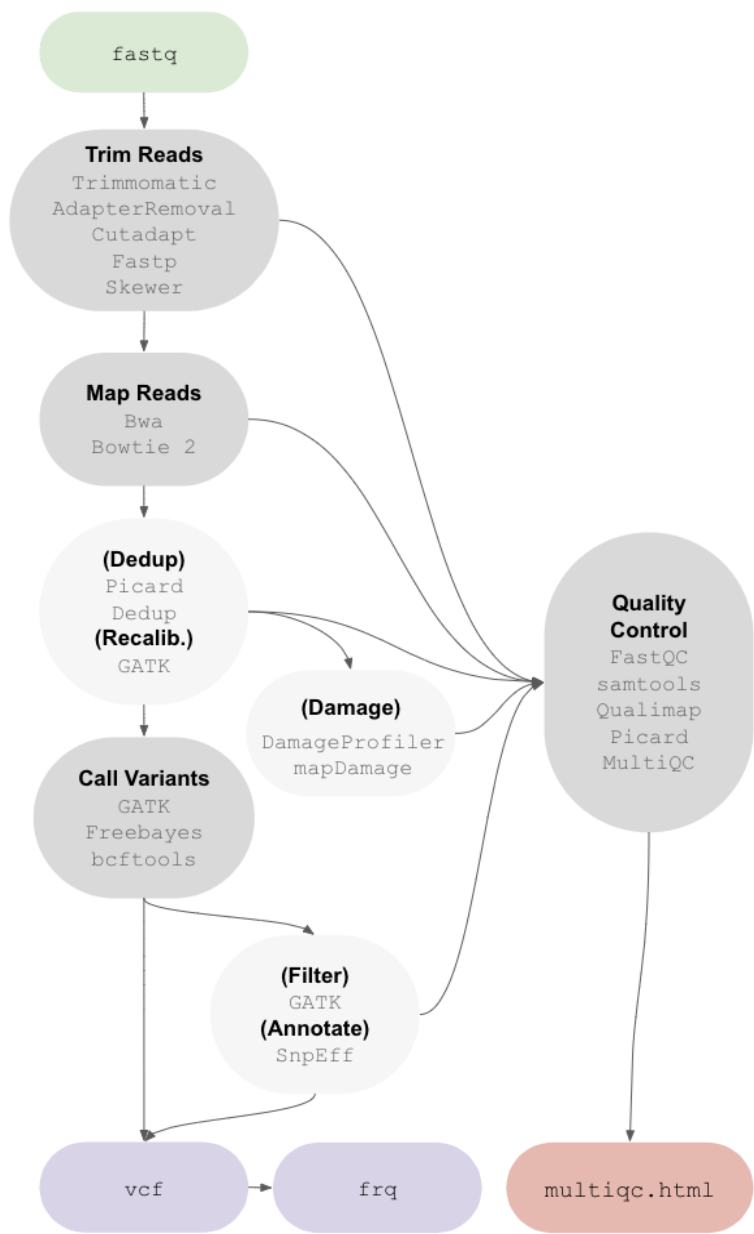

**Fig. 1 | g_renepipe workflow**
This high-level overview exemplifies the data flow of the g_renepipe pipeline, from raw sequencing data (fastq) to quality control report (multiqc.html), and final variant or frequency tables (vcf, frq). Minimal required steps are in dark gray, and optional data type-specific steps are in light gray. For detailed graphs and further details, visit the online wiki at:

github.com/moiexpositoalonsolab/grenepipe.

## Pipeline Steps

**Pre-run Reference Preparation**

Initially, the pipeline needs to prepare the reference genome to enable parallelization of the workload across genomic chromosomes or scaffolds in the reference sequence. In this step, we run `bwa index` [14,15], `samtools faidx` [15], `GATK CreateSequenceDictionary` (Genome Analysis Toolkit) [16], and `Tabix` [17], which create indices and other summaries of the reference genome.

### Read Trimming

Read trimming assists the subsequent read mapping by removing Illumina or other adapter sequences and low-quality bases from the reads. Typical read trimmers have two modes of operation, for single-end reads, and for paired-end reads. Some tools can furthermore merge paired end reads into a single read. We include the option to select from several tools for read trimming: `AdapterRemoval` [18,19], `Cutadapt` [20], `fastp` [21], `Skewer` [22], and `trimmomatic` [23]. The input and the output of this step are (compressed) `fastq.gz` files.

### Read Mapping

Next, the reads need to be aligned/mapped against the reference genome. The pipeline uses `bwa mem` [14] or `Bowtie2` [24] for this step, creating `sam`/`bam` files.

As an artifact of PCR-based DNA library preparation or sequencing chemistry and optical approaches, duplicated reads originating from a single fragment of DNA can occur in the data. We include a choice of two dedicated tools to mark such duplicates, namely, `Picard MarkDuplicates` [25,26] and `dedup` [25]. The pipeline also optionally allows to re-calibrate base quality scores using `GATK BaseRecalibrator` [16], which aims to detect systematic errors of the sequencing machine when estimating the accuracy of base calls. Both these steps are optional, and again produce `sam`/`bam` files.

### Ancient DNA and Damage Profiling

For input samples that are extracted from historical or ancient specimens, degradation and fragmentation of the DNA has to be studied. We include two optional tools for estimating damage patterns in ancient DNA samples: `mapDamage` [27,28] and the more recent `DamageProfiler` [29]. Their output is also included in the MultiQC quality control report.

### Quality Control

Quality control assesses the sequence data, e. g., to find issues in the library preparation and sequencing protocol. To this end, we include several dedicated quality control tools, namely `FastQC` [30], `samtools stats` [15] and `flagstat` [15], and `Qualimap` [31]. In addition, we run `Picard CollectMultipleMetrics` [26], which comprises the following tools: `AlignmentSummaryMetrics`, `BaseDistributionByCycle`, `GcBiasMetrics`, `InsertSizeMetrics`, `QualityByCycleMetrics`, `QualityScoreDistribution-Metrics`, and `QualityYieldMetrics`. Furthermore, several other tools described here such as read trimming and duplication removal, report statistics about their output. All these results are summarized by `MultiQC` [32], and compiled into an `html` report. The report allows researchers to quickly gain an overview of quality control statistics, but also to examine individual samples as needed.

### Variant Calling

The core step of the pipeline is to identify and retrieve ("call") genomic positions where one or more samples differ from the reference sequence, i. e., SNPs and short indels; see [33,34] for reviews and best practices. In this step, we currently implemented `GATK HaplotypeCaller` [16], `FreeBayes` [35,36], or `bcftools call` [37]. All tools are parallelized over contigs (i. e., typically, over chromosomes and/or scaffolds of the reference genome) for speedup. `GATK`

`HaplotypeCaller` does not call variants on the combined samples, and hence is further parallelized over individual samples. Optionally, an input `vcf` file can be provided to restrict the calling to known variants only [36].

Subsequently, we filter for SNPs and indels with `GATK SelectVariants`, and further allow to filter variants by their INFO and FORMAT annotations via `GATK VariantFiltration` or recalibrate variant qualities with `GATK VariantRecalibrator`. The output of this step is a gzip-compressed `vcf.gz` file that contains the genomic positions with variants for all samples. Optionally, genetic variants can be annotated with `SnpEff` [12] to obtain predictions of the effects of variants on genes, which also produces a `vcf` file.

**Frequency Calling**

Although $g_r$enepipe is agnostic to the type of genomic application (it can be used from SNP calling of individuals to discovery of somatic mutations), an important application for us is Pool-Seq, where DNA of individuals of a population has been combined ("pooled") in the same sequencing library. The resulting `vcf` typically contains diploid genotype calls (e.g. A/G). Our pipeline implements an output table with biallelic SNP positions in the rows, and allele frequencies of each library in the columns calculated based on allelic depth (REF/(REF+ALT)).

## Automation and distribution of jobs

A major advantage of $g_r$enepipe's Snakemake backend is its ability to automate workflows and to recover from failed compute jobs. Errors can occur in distributed cluster environments, e.g. brief network issues or broken hardware. By default, $g_r$enepipe tries each job three times, and aims to continue with independent jobs as long as possible. Even after persistent failure, once the researcher manually fixes the problem and restarts the run, the pipeline is able to continue seamlessly from the jobs that already succeeded.

In addition, we split data into chromosomes or scaffolds to improve data throughput (further splits would be possible), as these tasks are independent and can be run in parallel. Even some tasks on the same genomic fragments are also independent (for instance quality control and variant calling), so they are also run in parallel, maximizing computer usage.

## Example study and two-liner

Here and in the online wiki (see below) we showcase the use of $g_r$enepipe with a small toy dataset of 60k reads from pooled plant populations [38]. Once the sample names and file paths are provided in a table (`sample.tsv`) and referred to in the configuration file (`config.yaml`), the following two commands prepare the reference genome, and produce the final `vcf` from the example `fastq` files:

```
# https://github.com/moiexpositoalonsolab/grenepipe/wiki/Quick-Start-and-Full-Example

# Prepare the run, including pre-processing of reference genome
snakemake --use-conda --cores 4 --directory example/ --snakefile rules/prep.smk

# Produce vcf, frq, and MultiQC from example/samples/*.fastq
snakemake --use-conda --cores 4 --directory example/
```

We ran the same two commands on a larger dataset of 239 GB of compressed `fastq` files with 2.8 billion reads, which was distributed into 4,236 compute jobs and ran for ~40 hours beginning to end. This example was run on the Memex cluster (http://hpc.carnegiescience.edu) using the Slurm workload manager (https://slurm.schedmd.com), with a total availability of 88 nodes for pre-variant-calling steps and 10 nodes for variant calling and consolidation, with 24+ CPUs/Node, 128527+ MB Mem/Node.

## Outlook

g$_r$enepipe offers a streamlined and scalable workflow to process sequencing datasets to obtain single nucleotide variant and frequency calls, and extensive quality control statistics. In the future, we plan to extend g$_r$enepipe with additional tools popular in ecological & evolutionary genomics research, and extend workflows for allele frequency computations. We see this document and g$_r$enepipe as a living tool for students of bioinformatics that we will continue to expand and document to facilitate the adoption of genome sequencing technology in biology laboratories.

## Additional information

**Code Availability**

g$_r$enepipe is published under the GPL-3 license and is freely available at github.com/moiexpositoalonsolab/grenepipe. We appreciate user feedback, bug reports, and suggestions for additional tools.

**Author's Contributions**

LC and MEA conceived the project and wrote the manuscript. LC implemented and tested the g$_r$enepipe pipeline.

**Acknowledgments**

We would like to thank Moi Lab members for discussions of g$_r$enepipe, and Patricia Lang, Callie Rodgers Chappell, and Yunru Peng for their input on the project, their suggestions for tools and features to be added, and for their persistence as beta-testers. The soundtrack for this work was provided by Koji Kondo. This project was initiated as part of a collaborative network, Genomics of rapid Evolution to Novel Environments (GrENE-net.org), from where it inherits its name.

**Funding**

This work was supported by the Carnegie Institution for Science. The computing for this project was performed on the Memex cluster of the Carnegie Institution for Science.

**Disclosure statement**

The authors declare no competing interests. The funders had no role in study design, data collection and analysis, decision to publish, or preparation of the manuscript.